\documentclass{nature}
\usepackage[english]{babel}
\usepackage{tabulary,graphicx,amsmath,amsfonts,amssymb}
\usepackage{mathtools}
\usepackage{amsmath}
\usepackage[T1]{fontenc}
\usepackage{ucs}
\usepackage{tikz}
\usepackage{braket}
\usepackage{pgfplots}
\usepackage{csquotes}
\usepackage{hhline}

\usepackage{dcolumn}
\usepackage{tabularx}
\usepackage{longtable}
\usepackage{float}
\usepackage{epsfig}
\usepackage{latexsym}
\usepackage{amssymb}
\usepackage{amsmath}
\usepackage{amsthm}
\usepackage{amsfonts}
\usepackage{ifthen}
\usepackage{revsymb}
\usepackage{yfonts}
\usepackage{graphicx}
\usepackage{algpseudocode}
\usepackage{bbm}
\usepackage{multirow}
\usepackage{gensymb}
\usepackage{epstopdf}
\usepackage{listings}
\usepackage{color}
\definecolor{mygreen}{rgb}{0,0.6,0}
\definecolor{mygray}{rgb}{0.5,0.5,0.5}
\definecolor{mymauve}{rgb}{0.58,0,0.82}

\usepackage{afterpage}
\usepackage{verbatim}
\usepackage{soul}
\usepackage{lineno, blindtext}
\usepackage{footnote}
\usepackage{afterpage}
\usepackage{algpseudocode}
\usepackage[normalem]{ulem}
\usepackage{algorithm}
\usepackage{algpseudocode}
\usepackage{mathtools}

\usepackage{longtable}
\usepackage{xcolor}
\definecolor{darkred}  {rgb}{0.5,0,0}
\definecolor{darkblue} {rgb}{0,0,0.5}
\definecolor{darkgreen}{rgb}{0,0.5,0}
\usepackage{amsmath}

\newcommand{\ham}{\mathcal{H}}
\makeatletter
\renewenvironment{figure}
               {\@float{figure}}
               {\end@float}
\renewenvironment{figure*}
               {\@dblfloat{figure}}
               {\end@dblfloat}
\renewenvironment{table}
               {\@float{table}}
               {\end@float}
\renewenvironment{table*}
               {\@dblfloat{table}}
               {\end@dblfloat}

\makeatother

\usepackage{url,multirow,morefloats,floatflt,cancel,tfrupee}
\makeatletter

\AtBeginDocument{\@ifpackageloaded{textcomp}{}{\usepackage{textcomp}}}
\makeatother
\usepackage{colortbl}
\usepackage{xcolor}
\usepackage{pifont}
\usepackage[nointegrals]{wasysym}
\makeatletter

\def\mcWidth#1{\csname TY@F#1\endcsname+\tabcolsep}

\def\cAlignHack{\rightskip\@flushglue\leftskip\@flushglue\parindent\z@\parfillskip\z@skip}
\def\rAlignHack{\rightskip\z@skip\leftskip\@flushglue \parindent\z@\parfillskip\z@skip}

\if@twocolumn\usepackage{dblfloatfix}\fi 
\AtBeginDocument{
\expandafter\ifx\csname eqalign\endcsname\relax
\def\eqalign#1{\null\vcenter{\def\\{\cr}\openup\jot\m@th
  \ialign{\strut$\displaystyle{##}$\hfil&$\displaystyle{{}##}$\hfil
      \crcr#1\crcr}}\,}
\fi
}

\let\lt=<
\let\gt=>
\def\processVert{\ifmmode|\else\textbar\fi}

\@ifundefined{subparagraph}{
\def\subparagraph{\@startsection{paragraph}{5}{2\parindent}{0ex plus 0.1ex minus 0.1ex}%
{0ex}{\normalfont\small\itshape}}%
}{}

\newcommand\role[1]{\unskip}
\newcommand\aucollab[1]{\unskip}
  
\@ifundefined{tsGraphicsScaleX}{\gdef\tsGraphicsScaleX{1}}{}
\@ifundefined{tsGraphicsScaleY}{\gdef\tsGraphicsScaleY{.9}}{}
\def\checkGraphicsWidth{\ifdim\Gin@nat@width>\linewidth
	\tsGraphicsScaleX\linewidth\else\Gin@nat@width\fi}

\def\checkGraphicsHeight{\ifdim\Gin@nat@height>.9\textheight
	\tsGraphicsScaleY\textheight\else\Gin@nat@height\fi}

\def\fixFloatSize#1{}
\let\ts@includegraphics\includegraphics

\def\inlinegraphic[#1]#2{{\edef\@tempa{#1}\edef\baseline@shift{\ifx\@tempa\@empty0\else#1\fi}\edef\tempZ{\the\numexpr(\numexpr(\baseline@shift*\f@size/100))}\protect\raisebox{\tempZ pt}{\ts@includegraphics{#2}}}}

\AtBeginDocument{\def\includegraphics{\@ifnextchar[{\ts@includegraphics}{\ts@includegraphics[width=\checkGraphicsWidth,height=\checkGraphicsHeight,keepaspectratio]}}}

\def\URL#1#2{\@ifundefined{href}{#2}{\href{#1}{#2}}}

\def\UrlOrds{\do\*\do\-\do\~\do\'\do\"\do\-}%
\g@addto@macro{\UrlBreaks}{\UrlOrds}
\makeatother

\def\fixFloatSize#1{}

\newcolumntype{C}{>{\centering\arraybackslash}X}
  \pgfplotsset{compat=1.14}
\begin{document}
\setcounter{secnumdepth}{3}
\title{Spin-Boson Model to Demonstrate Quantum Tunneling in Biomolecules using IBM Quantum Computer}

\author{Yugojyoti Mohanta$^{1}$\thanks{E-mail: yugojyoti16@iiserbpr.ac.in}{ },
              Dhurjati Sai Abhishikth$^{2}$\thanks{E-mail: dabhishikth2717@iisertvm.ac.in}{ },
            Kuruva Pruthvi$^{2}$\thanks{E-mail: pruthvikuruva17@iisertvm.ac.in }{ },
              Vijay Kumar$^{3}$\thanks{E-mail: vk13ms149@iiserkol.ac.in}{ },
              Bikash K. Behera$^{3}$\thanks{E-mail: bkb13ms061@iiserkol.ac.in} { } \&
              Prasanta K. Panigrahi$^{3}$\thanks{Corresponding author.}\ \thanks{E-mail: pprasanta@iiserkol.ac.in}
              }
\maketitle 

\begin{affiliations}
 \item
    Department of Physical Sciences\unskip, 
    Indian Institute of Science Education and Research Berhampur\unskip, Ganjam\unskip, 760010\unskip, Odisha\unskip, India.
    \item
    Department of Physical Sciences\unskip, 
    Indian Institute of Science Education and Research Thiruvananthapuram\unskip, Vithura\unskip, 695551\unskip, Kerala\unskip, India.
    \item
      Department of Physical Sciences\unskip, 
    Indian Institute of Science Education and Research Kolkata\unskip, Mohanpur\unskip, 741246\unskip, West Bengal\unskip, India.
\end{affiliations}
        
\begin{abstract}
Efficient simulation of quantum mechanical problems can be performed in a quantum computer where the interactions of qubits lead to the realization of various problems possessing quantum nature. Spin-Boson Model (SBM) is one of the striking models in quantum physics that enables to describe the dynamics of most of the two-level quantum systems through the bath of harmonic oscillators. Here we simulate the SBM and illustrate its applications in a biological system by designing appropriate quantum circuits for the Hamiltonian of photosynthetic reaction centers in IBM's 5-qubit quantum computer. We consider both two-level and four-level biomolecular quantum systems to observe the effect of quantum tunnelling in the reaction dynamics. We study the behaviour of tunneling by changing different parameters in the Hamiltonian of the system. The results of SBM can be applied to various two-, four- and multi-level quantum systems explicating electron transfer process.
\end{abstract}

\section{Introduction \label{qsb_Introduction}}
Quantum computers have proved their superior power over classical ones while solving certain problems \cite{qsb_GroverPRL1997,qsb_DeutschPRSLA1992,qsb_FeynmanIJTP1982,qsb_ChildsPASTC2003,qsb_SimonFCSPAS1994,qsb_GerjuoyAJP2005} among which simulation of the quantum systems has been a special attraction due to the exponential improvement in speeds and computational resources. Efficient simulation of quantum systems \cite{qsb_AspuruScience2005,qsb_FarhiScience2001,qsb_HanCEC2000,qsb_ChuangNature1998,qsb_JonesNature1998,qsb_GuldeNature2003,qsb_HarrowPRL2009,qsb_GerritsmaNat2010} has played a major role in motivating scientists coming up with the idea of a quantum computer \cite{qsb_FeynmanIJTP1982}. The field of simulation of quantum mechanical problems in a quantum computer is progressing very fast and it has applications in many scientific branches like many-body theory \cite{qsb_TsengPRA1999,qsb_NegrevergnePRA2005,qsb_PengPRL2009,qsb_FengPRL2013}, condensed matter \cite{qsb_EdwardsPRB2010,qsb_ZhangPRA2009}, spin models \cite{qsb_GarciaPRL2004,qsb_LanyonSci2011}, quantum phase transitions \cite{qsb_GreinerNat2002,qsb_PolletPRL2010}, quantum chemistry \cite{qsb_LidarPRE1999}, quantum chaos \cite{qsb_WiensteinPRL2002}, interferometry \cite{qsb_LeibfriedPRL2002,qsb_ViyuelanpjQI2018,qsb_LangfordNatComm2017} and so on. Simulation by a quantum computer has been found to be more effective than simulated it in a classical computer \cite{qsb_BulutaScience2009}.

One of the most fundamental models of open quantum systems is the Spin-Boson Model (SBM) \cite{qsb_MascherpaPRL2017,qsb_WeissWS1999}, which comprises a two-level system and a large number of quantum harmonic oscillators linearly coupled to it and acting as the environment \cite{qsb_LeppakangasPRA2018}. The influence of these degrees of freedom on the dynamics of the spin can be computed from the strength of the couplings between the spin, each oscillating mode and the frequency of the modes \cite{qsb_SchultenCP1991,qsb_NonellaJPC1991,qsb_KrishtalikBBAB1995,qsb_XuChemPhys1994,qsb_DeVaultCUP1980}. Though exact solution of the spin-boson model is not possible \cite{qsb_ReinholdPRB1994}, various simulation techniques such as real time quantum Monte Carlo simulations \cite{qsb_LuckPRE1998} and exact Monte Carlo simulations \cite{qsb_LeePRE2012} and approximation methods such as Noninteracting-blip approximation \cite{qsb_LeggettRMP1987} and Bloch-Redfield equation have been extensively used to study the dynamics of this model. While most of the simulation performed in classical systems neglects the tunneling part \cite{qsb_XuChemPhys1994}, through simulation of SBM on a quantum computer, we can obtain insightful results by incorporating the tunneling part in the Hamiltonian of the quantum system.              

Here we apply the SBM to explain the dynamics of biomolecules in the photosynthetic reaction center, i.e. electron transfer reaction in Rhodopseudomonas viridis \cite{qsb_XuChemPhys1994,qsb_KiramaierPR1987,qsb_HuPhysToday1997} and further extend this model to a four-level system. We study the time evolution of the given system from reactant state to product and from product to reactant state for different tunnelling parameters in such biological system at normal room temperature. We use IBM's quantum computer, `IBM 5 Tenerife' (ibmqx4) to simulate the Hamiltonian of this systems, where many research problems have already been tackled \cite{qsb_GarciaarXiv2017,qsb_DasarXiv2017,qsb_BeheraQIP2017,qsb_Hegadearxiv17,qsb_MajumderarXiv2017,qsb_SisodiaQIP2017,qsb_DasharXiv2017,qsb_WoottonQST2017,qsb_BertaNJP2016,qsb_DeffnerHel2017,qsb_HuffmanPRA2017,qsb_AlsinaPRA2016,qsb_YalcinkayaPRA2017,qsb_GhosharXiv2018,qsb_KandalaNAT2017,qsb_RodriguezarXiv2017,qsb_SchuldEPL2017,qsb_SisodiaPLA2017,qsb_TannuarXiv2018,qsb_WoottonarXiv2018,qsb_HarperarXiv2018,qsb_4BKB6arXiv2018,qsb_5BKB6arXiv2018,qsb_6BKB6arXiv2018,qsb_Roffe2018,qsb_PlesaAJGR2018,qsb_ManabputraarXiv2018,qsb_JhaarXiv2018,qsb_GangopadhyayQIP2018,qsb_AggarwalarXiv2018}. The architecture of the 5-qubit chip `ibmqx4' is discussed in Methods Section \ref{qsb_Methods}.

\section{Results \label{qsb_Results}}
\subsection{Two-Level System} The Hamiltonian of Spin-Boson Model for a two-level system is given by \cite{qsb_LeppakangasPRA2018,qsb_LeggettRMP1987}, 

\begin{equation}\label{qsb_Eq1}
\ham=-\frac{\hbar \Delta  \hat{\sigma_x}}{2}+\frac{ \epsilon \hat{\sigma_z}}{2}+\frac{q  \hat{\sigma_x} \Sigma c_\alpha x_\alpha}{2\hbar}+\Sigma \hbar w_\alpha \hat{b^\dagger_\alpha} \hat{b_\alpha}
\end{equation}
where $\hat{\sigma_x}$ and $\hat{\sigma_z}$ are Pauli operators, $\epsilon$ represents the energy difference between two states, $\hbar$ represents the reduced Plank's constant, $\Delta$ represents the hopping rate, q represents the protein conformation for initial state and $c_\alpha$ describes the strength of the coupling of the electron transfer to the $\alpha$ th oscillator and $\hat{b^ \dagger},\hat{b}$ are bosonic creation and annihilation operators and tunnelling parameter is given by $\frac{\hbar\Delta}{2}$.
        
\begin{figure}[H]
\centering
\includegraphics[width=\linewidth]{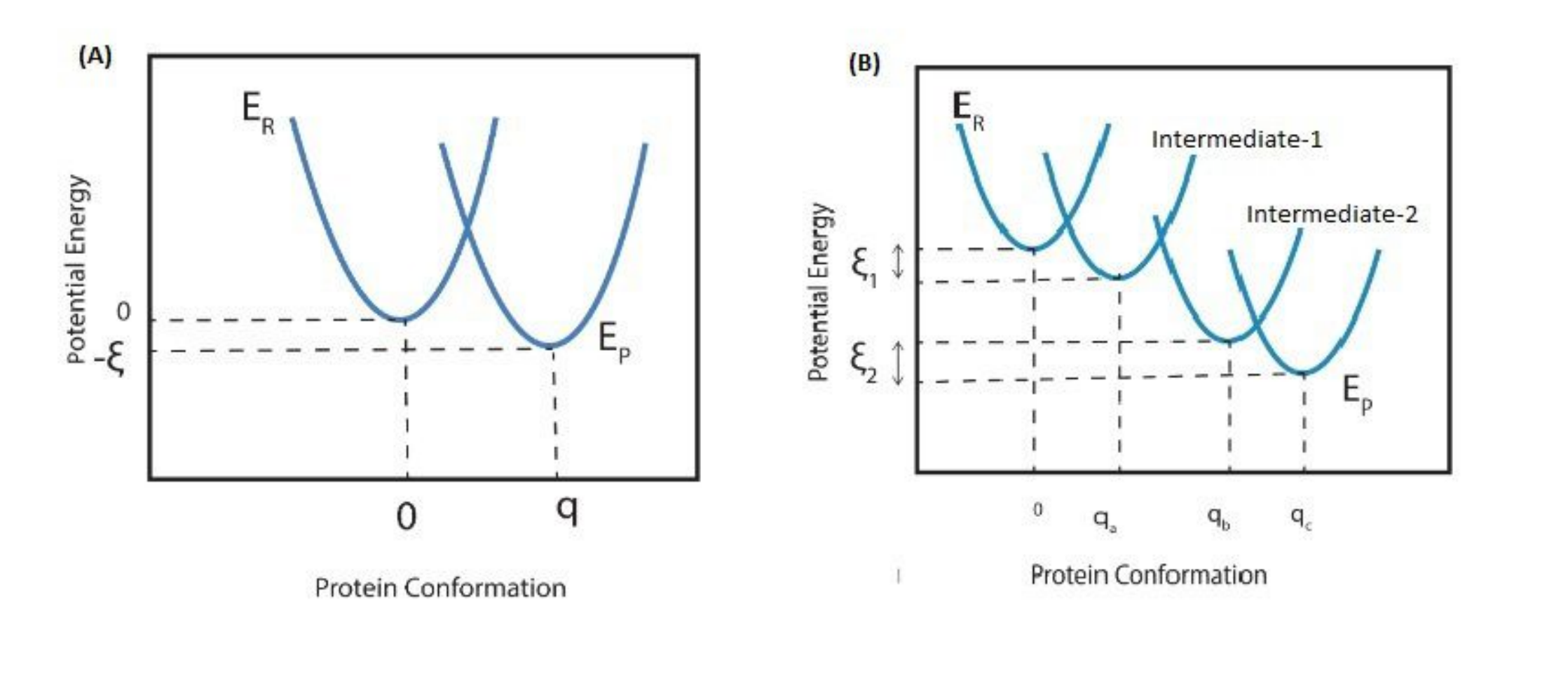}
\caption{\textbf{Marcus diagram for two-level and four-level systems.} \textbf{Case A}: $E_{R}$ and $E_{P}$ represent the energies of reactant and product respectively, $\epsilon$ is the energy difference between reactant and product, and q is the protein conformation of the product. \textbf{Case B}: $E_{R}$ and $E_{P}$ represent the energies of reactant and product respectively. Intermediate-1 and Intermediate-2 are the two intermediate states of the reaction. $\epsilon_{1}$ is the energy difference between reactant and intermediate-1 state and $\epsilon_{2}$ is the energy difference between intermediate-2 and the product state. $q_{a}$, $q_{b}$ and $q_{c}$ are the protein conformation of intermediate-1, intermediate-2 and the product respectively.}
\label{qsb_Fig1}
\end{figure}

Here we apply the Spin-Boson (SB) Hamiltonian to a two-level biological system which is similar to Rhodopseudomonas viridis's  photosynthetic reaction center \cite{qsb_XuChemPhys1994,qsb_WarrenCCR2013}. The system can be visualised using a Marcus energy diagram (Fig. \ref{qsb_Fig1} \textbf{A}) which consists of two potential wells with the energy difference $\epsilon$. The distance between the two potentials is denoted as q which is so called protein conformation. $E_{R}$ and $E_{P}$ are the two energies of the reactant and product states, where  $\epsilon=E_{R}-E_{P}$. 

\subsection{Four-Level System}

The Spin-Boson Hamiltonian can be extended to a four-level system which describes electron transfer methods in four-level photosynthetic reaction. This is done by adding two intermediate energy states in between reactant state and product state. The system can be represented using a Marcus energy diagram Fig. as illustrated in Fig. \ref{qsb_Fig1} \textbf{B}. To simplify the system we model the Hamiltonian in such a way that tunnelling is only allowed in between reactant state ($E_R$) and Intermediate-1 and in between Intermediate-2 and product state ($E_P$). $\epsilon_1$ and $\epsilon_2$ denote the energy differences between the reactant and Intermediate-1, and the Intermediate-2 and product state respectively.  
 
The Hamiltonian describing the four-level system is given by \cite{qsb_GilmoreJPCM2005,qsb_LeppakangasPRA2018},
 
\begin{equation} \label{qsb_Eq2}
\ham= - \frac{\hbar \Delta_1  \hat{\sigma_x ^1}}{2}+\frac{ \epsilon_1 \hat{\sigma_z ^1}}{2}+\frac{q  \hat{\sigma_x ^1} \Sigma c_\alpha x_\alpha}{2\hbar}+\Sigma \hbar w_\alpha\hat{a^\dagger _\alpha} \hat{a_\alpha}
-\frac{\hbar \Delta_2  \hat{\sigma_x ^2}}{2}+\frac{ \epsilon_2 \hat{\sigma_z ^2}}{2}+\frac{q  \hat{\sigma_x ^2} \Sigma c_\beta x_\beta}{2\hbar}+\Sigma \hbar w_\beta \hat{b^ \dagger _\beta} \hat{b_\beta}+J_{12}(\hat{\sigma_x ^1} \hat{\sigma_x ^2} +\hat{\sigma_y ^1} \hat{\sigma_y ^2})
\end{equation}

where $\hat{\sigma_x^i}$ and $\hat{\sigma_z^i}$ are the Pauli operators acting on the $i$th qubit, $\epsilon_1$ and $\epsilon_2$ are the energy differences between the states, $\hbar$ is the reduced Plank's constant, $\Delta_1$ and  $\Delta_2$ represent the hopping rates between two energy states, q is the distance between two wells, and $c_\alpha$ and $c_\beta$ describe the strength of the coupling of the electron transfer to the $\alpha$th and $\beta$th oscillator respectively. Here $\hat{a^\dagger}$ and $\hat{b^\dagger}$ are the bosonic creation operators, $\hat{a}$ and $\hat{b}$ are the annihilation operators, and $J_{12}$ is the coupling constant between the two qubits \cite{qsb_GilmoreJPCM2005,qsb_HussainSciJPhys2012,qsb_DacresCodChemRev2013}.
 
The reaction dynamics of photosynthetic reaction center (i.e., the conversion of reactant to product state) and the tunnelling effect \cite{qsb_HopfieldPNASU1974} in two-level and four-level system is studied by making appropriate quantum circuit. The Hamiltonian for the system is simulated using IBM'S five-qubit quantum computer for different instances of time (t). To understand the quantum tunnelling efficiently, we took the values of hopping rates ($\Delta_1$, $\Delta_2$), energy differences ($\epsilon_1$, $\epsilon_2$) to be equal for both of the qubits in the four-level system and took approximate values for the terms containing summation. The executions and graphs are shown in Results Section \ref{qsb_Results}.

\subsection{Quantum Circuits}
\begin{figure}[H]
\centering
\includegraphics[width=\linewidth]{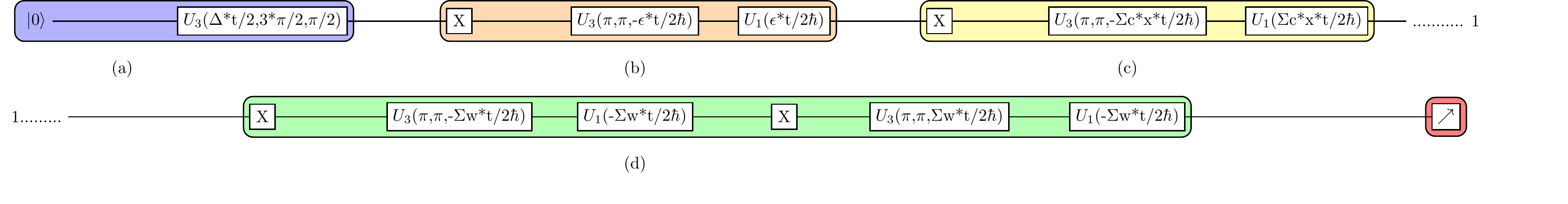}
\caption{\textbf{Circuit diagram for two-level system Hamiltonian.} \textbf{(a)} represents the tunneling term ($\frac{\hbar \Delta} {2}$), \textbf{(b)} represents the energy difference ($\epsilon$), \textbf{(c)} represents the coupling between system and bath and \textbf{(d)} represents the bath term.}
\label{qsb_Fig2}
\end{figure}

\begin{figure}[H]
\centering
\includegraphics[width=\linewidth]{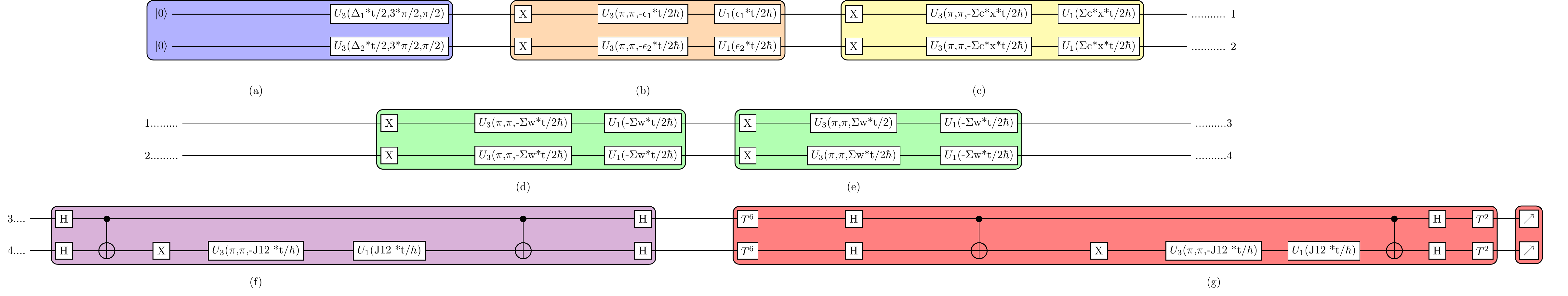}
\caption{\textbf{Circuit diagram for four-level system Hamiltonian.} \textbf{(a)} represents the tunnelling terms ($\frac{\hbar \Delta_1}{2}$) and ($\frac{\hbar \Delta_2}{2}$). \textbf{(b)} represents the energy difference between two states $\epsilon_1$ and $\epsilon_2$. \textbf{(c)} represents the coupling with the bath, \textbf{(d) \& (e)} represent the bath term for each qubit respectively, \textbf{(e) \& (f)} represent the coupling term between 2-qubits.}
\label{qsb_Fig3}
\end{figure}

For the two-level and the four-level system, the design of the circuit of the SB Hamiltonian is given Figs. \ref{qsb_Fig2} \& \ref{qsb_Fig3}. The values of all the parameters in the Hamiltonian are discussed in Methods Section \ref{qsb_Methods}. The Hamiltonian is executed in quantum computer and the formation of product of the reaction is observed and the probability of formation is plotted against time for different tunnelling matrix elements. We observed changes in electron transfer rate as the tunnelling parameter or the value of tunneling matrix element varies in the Hamiltonian. The tunnelling effect can be observed as the reactant forming the product and vice-versa. We plotted a bar graph for time evolution containing states of both the systems (two-level and four-level) with their probabilities. The quantum circuits for two-level are shown in Fig. \ref{qsb_Fig2}.
 
\begin{figure}[H]
\centering
\includegraphics[width=\linewidth]{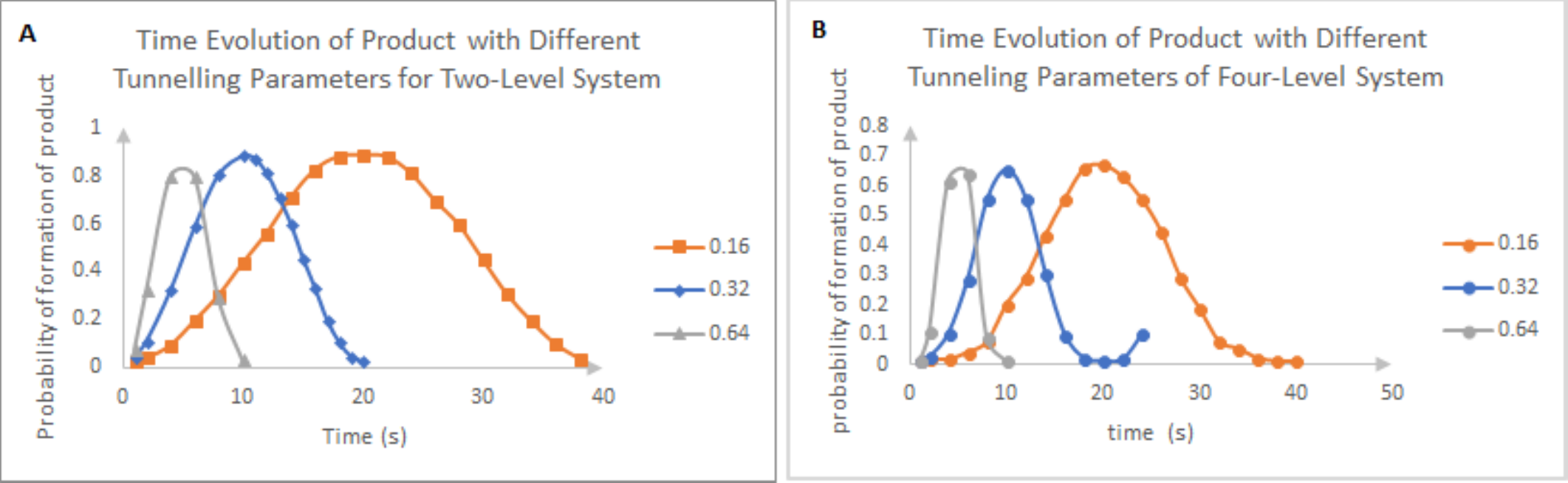}
\caption{\textbf{Time evolution of product of two-level and four-level system.} {Case A}: Time evolution of product with change in tunneling matrix ($\frac{\hbar\Delta}{2}$) of two-level system describing photosynthetic reaction center. We evolve the system by  increasing time till significant tunnelling is observed from the reactant state to the product state and vice-versa. The graphs were plotted for three different tunnelling parameters (orange graph is for $0.16 \times 10^{-22}J$, blue for $0.32\times10^{-22}J$ and grey for  $0.64\times10^{-22}J$) and time shown in graph is in order of $10^{-12}$s. We can clearly infer that the time taken for tunnelling decreases or the tunnelling rate increases as the tunnelling parameter increases. Initially, the product state $\ket{1}$ has 0 probability and reactant state $\ket{0}$ have probability 1, then slowly as time evolves the probability of formation of the product reaches 1 and reactant state 0. After further time evolution, the products get tunnel back to the reactant state which is referred as tunneling cycle. \textbf{Case B}: For two qubit system the product state ($\ket{11}$), intermediate states ($\ket{01}$) and ($\ket{10}$) have 0 probability and reactant state ($\ket{00}$) have probability 1, then slowly as time evolves the probability of formation of the product reaches 1 and reactant state probability falls to 0.}
\label{qsb_Fig4}
\end{figure}

Tunnelling of reaction state into product state and vice-versa can be visualised in Fig. \ref{qsb_Fig4} as the time (t) evolves. It is observed that the time taken for tunneling from reaction state to product state is approximately decreases by half as the tunnelling parameter doubles.

the initial configuration of the two-level system illustrating electron transfer in photosynthetic reaction center is taken such that reactant state has a probability of 1 and product state has a probability of 0. As the time evolves in the system, concentration of product increases and concentration of reactant decreases. For tunnelling parameter 0.16 $\times$ $10^{-22}J$, it reaches nearly equilibrium stage (i.e., the concentration of product state and reaction state are equal) at time t=$11 \times 10^{-12}$s. On further evolving the system, the probability of product state nearly reaches 1 and reactant state reaches 0, at about t=$22 \times 10^{-12}$s. After then, the system is further evolved and the tunnelling takes place in the reverse direction, i.e., the product gets tunnel back to the reactant state. This can be considered as tunnelling cycle and the systems with tunnelling parameters $0.16 \times 10^{-22}$J, $0.32 \times 10^{-22}$J and $0.64 \times 10^{-22}J$ take about $40 \times 10^{-12}$s, $20 \times 10^{-12}$s, $10 \times 10^{-12}$s respectively for one full tunnelling cycle.
 
In the four-level system describing photosynthetic reaction center, initially, the reactant has a probability of 1 and Intermediate-1, Intermediate-2, and the product state have a probability 0. As we evolve the system with time, the reactant state gets tunnelled into Intermediate-1 state ($\ket{01}$), and as the reaction progresses $\ket{01}$ gets converted to $\ket{10}$, and $\ket{10}$ state tunnels in the product state ($\ket{11}$). For the system with tunnelling parameter $0.16 \times 10^{-22}J$, we reach equilibrium at t=$14\times10^{-12}$. At $t=22\times10^{-12}$s, the product state has a probability of 1, and all other states have probability 0 and the system returns to the initial condition. This phenomena can be here referred as tunneling cycle as the reactant state tunnels to the product state and the product state then tunnels back to the reactant state.

\begin{figure}[H]
\centering
\includegraphics[width=\linewidth]{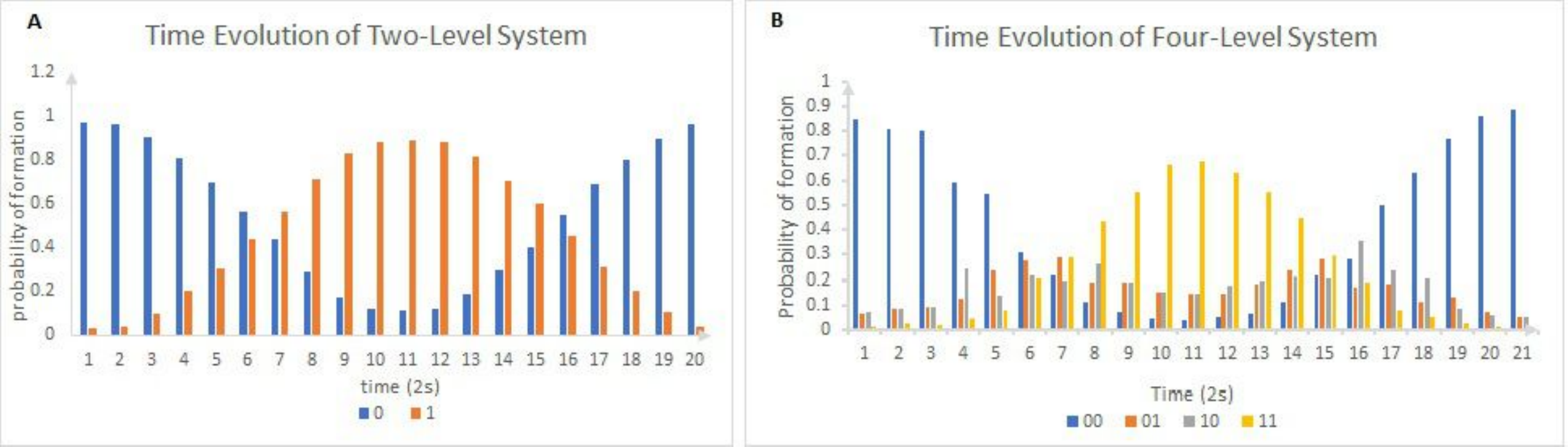}
\caption{\textbf{Bar graph for time evolution of reactants, products and intermediate states.} {Case A} : Bar graph showing the probability of formation of reactant and product of a two-level system as the time evolves at tunnelling parameter = $0.16\times10^{-22}J$. $\ket{0}$ state is the reactant state shown in blue and $\ket{1}$ state is the product state in orange. \textbf{Case B}: Bar graph showing the probability of formation of reactant and product of a four-level system as the time evolves at tunnelling parameter = $0.16\times10^{-22}J$. The four colours blue, orange, grey and yellow represent the reactant ($\ket{00}$), Intermediate-1 ($\ket{01}$), Intermediate-2 ($\ket{10}$) and the product ($\ket{11}$) state respectively. In both \textbf{Cases A} and \textbf{B} the time taken is in order of $10^{-12}$s.}
\label{qsb_Fig5}
\end{figure}

\begin{figure}[H]
\centering
\includegraphics[width=\linewidth]{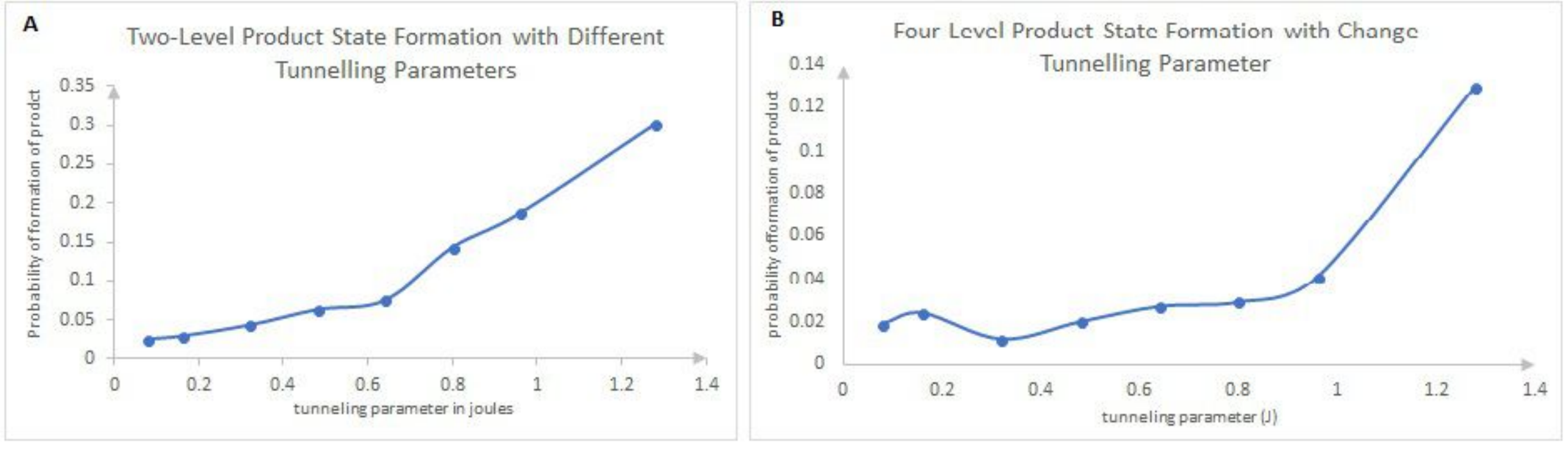}
\caption{\textbf{Product formation with different tunnelling parameter.} {\textbf{Case A}: Graph showing the changes in concentration of product when there is a change in tunnelling parameter at t= $10^{-12}$s for the two-level system. We can observe that the concentration of product significantly increases with increase in tunneling parameter at a constant time. \textbf{Case B}: Graph showing the changes in the concentration of product when there is a change in tunnelling parameter at t= $10^{-12}$s for four-level system. In both cases tunneling parameter is of order the  $10^{-22}$}.}
\label{qsb_Fig6}
\end{figure}

Fig. \ref{qsb_Fig6} gives a grasp on how tunnelling parameter affects the formation of the product, and we can see that the reactant is getting tunnelled into the product at t=$10^{-12}$s as the tunnelling parameter ($\frac{\hbar\Delta}{2})$ changes.
.

\section{Discussion \label{qsb_Discussion}}
We have successfully demonstrated the effect of quantum tunnelling  and studied the reaction dynamics of the biological system at normal room temperature consisting of two-level system and four-level system using spin-boson model. We studied how the system evolves for different tunneling parameters and found that rates of conversion of reactant to product increase with increase in tunneling parameter values. Study of four level system can be improved by allowing the tunnelling in between all four states. This can be further analyzed by taking different tunneling values for each qubits and different coupling values in between qubits. System dynamics can be observed at different temperature conditions and at different reaction rates in the system. This model can be extended for more than four state electron transfer reaction consisting of multi-intermediate states. Similar model can be applied to simulate the dynamics of Fluorescence Resonance Energy Transfer \cite{qsb_GilmoreJPCM2005,qsb_HussainSciJPhys2012}, Retinal in Rhodopsin environment and a quantum dot in a polar solvent \cite{qsb_GilmoreJPCM2005}. The spin-boson model can be used for wide range of physical processes including electron transfer \cite{qsb_LeggettRMP1987,qsb_MarcusBBARB1985,qsb_WangNJP2008}, macroscopic quantum coherence \cite{qsb_WeissJLTP1987} and hydrogen tunneling \cite{qsb_SuarezJCP1991}.      

\section{Methods} \label{qsb_Methods}

The values taken for simulating the above Hamiltonians are fixed for normal room temperature. The values of the following parameters are, tunnelling parameter ($\frac{\hbar\Delta} {2}$)= 3.2 $\times$ $10^{-23}J$, q = 1 \cite{qsb_XuChemPhys1994}, energy difference ($\epsilon) = 4.8 \times 10^{-20}J$, $\Sigma c_\alpha x_\alpha \approx 7.484 \times 10^{-16}J$ \cite{qsb_XuChemPhys1994} and $\Sigma\omega_\alpha$ is $1.1594 \times 10^{-16}J$ and number of harmonic oscillators \cite{qsb_XuChemPhys1994} constitute the bath are taken as order of $ 10^{4}$. For the four-level system the above same values are taken for both the qubits and except the value of $J_{12}$ which is taken as $2.603 \times 10^{-16}J$ \cite{qsb_HussainSciJPhys2012,qsb_DacresCodChemRev2013}. 

\textbf{Simulation of Hamiltonian}:
For simulation of the Hamiltonian \cite{qsb_NielsenCUP2002,qsb_HegadearXiv2017,qsb_LlyodSci1996} we use first order Trotter decomposition, which is given by,
\begin{equation}
e^{-i\ham t} = e^{-i \ham_1 t}e^{-i \ham_2 t}...e^{-i \ham_n t} +O\big(t^2)
\end{equation}
where $\ham_1, \ham_2,...,\ham_n$ are Hamiltonians acting on local subsystems involving k-qubits of an n-qubit system. The system Hamiltonian can be written as, $\ham = \sum_{1} ^ {n} \ham_k$. Then the Hamiltonian is  decomposed into a sequence of unitary transformations which can be implemented by using any set of universal quantum gates. In the above model,the Hamiltonian for a two state system is given by Eq. \eqref{qsb_Eq1}.

To implement the Trotter decomposition, we use  $\ham = \ham_1 + \ham_2 + \ham_3 +\ham_4$. Time evolution of quantum mechanical system is given by unitary transformation $\hat{U(t)}$, where
\begin{equation}
\hat{U(t)}=  e^{\frac{-i \ham t}  {\hbar}}
\end{equation}

Time evolution of $\ham_1$ is given by 
\begin{equation}
e^{\frac{-i \ham t}  {\hbar}}  =  e^{\frac{i \Delta  \hat{\sigma_x}t}{2} }
\end{equation} 
where $\ham_1= - \frac{\hbar \Delta  \hat{\sigma_x }}{2}$.
which can be written in matrix form 
$
\begin{bmatrix}\label{m1}
    \cos{\frac{ \Delta  t}{2}} & \iota\sin{\frac{ \Delta  t}{2}}\\
    \iota\sin{\frac{ \Delta  t}{2}} & \cos{\frac{ \Delta  t}{2}} 
\end{bmatrix}
$

The above matrix acts on a single qubit. Here t is the time elapsed since the beginning of the experiment. As the IBM Q Experience is a static system, by taking t as a controllable parameter we are able to effectively simulate the dynamics and time evolution of a two state biological system .

This matrix can be implemented on IBM Quantum Experience by using IBM's  $U3(\theta,\lambda,\phi )$. This matrix  can written in term of $U3(\theta,\lambda,\phi )$ by setting the parameters in the form U3$(\frac{ \Delta t}{2},\frac{ 3\pi}{2},\frac{ \pi}{2} )$.

Similarly, $\ham_2$, $\ham_3$ and $\ham_4$ can be implemented and quantum circuit for two-level system can be designed by using  $U3(\theta,\lambda,\phi)$, $U1(\theta)$ and X Gates as shown Fig. \ref{qsb_Fig2}.

For four level system, which is given by Hamiltonian Eq. \ref{qsb_Eq2} and $\ham$ can be written as,

$\ham = \ham_1 + \ham_2 + \ham_3 +\ham_4+\ham_5+\ham_6+\ham_7+\ham_8+\ham_9$

and $\ham_1$ takes the matrix form
$
\begin{bmatrix}\label{m2}
    \cos{\frac{ \Delta_1  t}{2}} & \iota\sin{\frac{ \Delta_1 t}{2}}\\
    \iota\sin{\frac{ \Delta_1 t}{2}} & \cos{\frac{ \Delta_1 t}{2}} 
\end{bmatrix}
$
acting on the qubit 1 and identity on qubit 2. 

Similarly, $ \ham_2 , \ham_3 ,\ham_4,\ham_5,\ham_6,\ham_7,\ham_8 \& \ham_9$ can be implemented and quantum circuit for four level system can be designed by using $U3(\theta,\lambda,\phi )$, $U1(\theta )$, X, H, T and CNOT Gates \cite{qsb_RaeisiNJP2012}, which is shown in Fig. \ref{qsb_Fig3}.

\begin{figure}[H]
\centering
\includegraphics[scale=2]{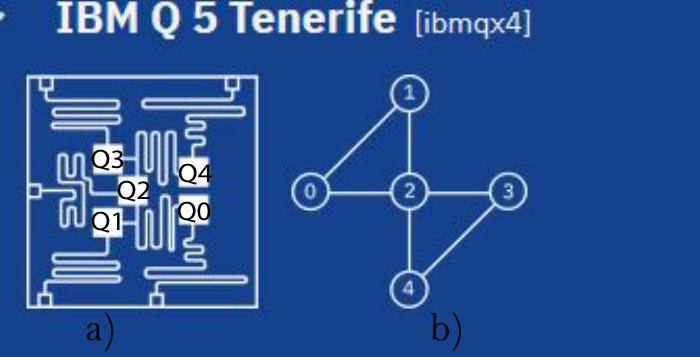}
    \caption{\textbf{Architecture of IBM Q5}\textbf{(a)} A picture shows the chip layout of 5-qubit quantum processor Tenerife [ibmqx4]. All 5 transmon qubits (Q0,Q1,Q2,Q3 and Q4) are connected with the two coplanar waveguide (CPW) resonators as shown. Qubits Q2, Q3 and Q4 are coupled by coplanar waveguide (CPW) resonators with resonances around 6.6 GHz, Qubits Q0, Q1 and Q2 are coupled by another coplanar waveguide (CPW) resonators with resonances around 7.0 GHz. Each qubit has a dedicated CPW for the control and readout. Two qubits gates are connected with a superconducting bus resonator. \textbf{(b)} The IBM Q experience uses the cross-resonance interaction as the basis for the CX-gate which is as follows: $\{Q1 \rightarrow (Q0), Q2\rightarrow Q0, Q1, Q4), Q3 \rightarrow (Q2, Q4)\}$, where $i \rightarrow (j)$ means $i$ and $j$ denote the control qubit and the target qubit respectively for implementation of CNOT gate in the chip. }
\label{qsb_Fig7.pdf}
\end{figure}

\textbf{Experimental Architecture} 
The experimental device specification of IBM Q 5 Tenerife [ibmqx4] chip are shown in Table \ref{qsr_sup_tab1} \cite{qsb_github}, the Fig. \ref{qsb_Fig7.pdf} depicts the connection and control of five qubits (Q0, Q1, Q2, Q3 and Q4).
The single-qubit gate error is of the order $10^{-3}$. The multi-qubit and readout error are of the order $10^{-2}$. Randomized benchmarking is used to measure the gate errors.

\begin{table}[H]
\centering
\begin{tabular}{ c c c c c c }
\hline
\hline
Qubits & $T^{||}_{1}$ ($\mu s$) & $T^{\perp}_{2}$ ($\mu s$) & GE$^{\dagger}$ & RE$^{\ddagger}$ \\
\hline
Q0 & 50.80 & 13.90 & 0.77 & 6.40 \\
Q1 & 56.30 & 57.70 & 1.63 & 6.10 \\
Q2 & 42.00 & 49.90 & 1.20 & 6.00 \\ 
Q3 & 33.10 & 15.30 & 3.01 & 11.00 \\
Q4 & 52.30 & 26.20 & 0.94 & 5.68 \\
\hline
\hline
\end{tabular}\\
$||$ Relaxation time, $\perp$ Coherence time, $\dagger$ Gate Error, $\ddagger$ Readout Error \\
\caption{\textbf{Experimental parameters of the device `ibmqx4' Tenerife.}}
\label{qsr_sup_tab1}
\end{table}

\bibliographystyle{}

\section*{Acknowledgements} The authors acknowledge Avinash Dash for useful dicussion; Daattavya Aggarwal and Deepankar Sarmah for helping in using QISKit. Y.M., D.S.A. and K.P. acknowledges the hospitality of Indian Institute of Science Education and Research Kolkata during the project work. Y.M. and B.K.B. acknowledge financial support of Inspire fellowship provided by Department of Science and Technology (DST), Govt. of India. We acknowledge the support of IBM Quantum Experience for providing access to the IBM quantum processors. The views expressed are those of the authors and do not reflect the official position of IBM or the IBM Quantum Experience team. 

\section*{Author contributions}
The idea of simulating biological system by Spin-Boson Model was initially proposed by V.K. Theoretical analysis, design of quantum circuit simulation, collection and analysis of data were performed by Y.M., D.S.A. and K.P. Y.M., D.S.A., K.P., V.K. and B.K.B. contributed to the composition of the manuscript. All authors have given approval to its final version. B.K.B. supervised the project. Y.M., D.S.A., K.P., V.K. and B.K.B. have completed the project under the guidance of P.K.P.   

\section*{Competing interests}
The authors declare no competing financial interests. Readers are welcome to comment on the online version of the paper. Correspondence and requests for materials should be addressed to P.K.P. (pprasanta@iiserkol.ac.in).  
  
\end{document}